\documentclass[
reprint,
superscriptaddress,
showpacs,preprintnumbers,
amsmath,amssymb,
aps,
prb,
floats,
]{revtex4-1}

\usepackage{graphics}
\usepackage{subfigure}

\draft 

\begin{document}




\preprint{}

\def\kfs122{KFe$_2$Se$_2$}
\def\kfsx22{K$_{1-x}$Fe$_2$Se$_2$}
\def\lofa{LaFeAsO}
\def\loffa{LaFeAsO$_{1-x}$F$_x$}
\def\bfa122{BaFe$_2$As$_2$}
\def\fs11{FeSe}

\title{Electronic Structure of \kfs122 from First-Principles Calculations} 

\author{Chao Cao}
  \affiliation{Condensed Matter Physics Group,
  Department of Physics, Hangzhou Normal University, Hangzhou 310036, China}

\author{Jianhui Dai}
  \affiliation{Condensed Matter Physics Group,
  Department of Physics, Hangzhou Normal University, Hangzhou 310036, China}
  \affiliation{Zhejiang Institute of Modern Physics, Department of Physics, Zhejiang University, Hangzhou 310027, China}





\date{\today}

\begin{abstract}
Electronic structure and magnetic properties for iron-selenide
\kfs122 are studied by first-principles calculations. The ground
state is collinear antiferromagnetic with calculated 2.26 $\mu_B$
magnetic moment on Fe atoms; and the $J_1$, $J_2$ coupling strengths
are calculated to be 0.038 eV and 0.029 eV. The states around $E_F$
are dominated by the Fe-3d orbitals which hybridize noticeably to
the Se-4p orbitals. While the band structure of \kfs122 is similar
to a heavily electron-doped \bfa122 or \fs11 system, the Fermi
surface of \kfs122 is much closer to \fs11 system since the electron
sheets around $M$ is symmetric with respect to $x$-$y$ exchange.
These features, as well as the absence of Fermi surface nesting,
suggest that the parent \kfs122 could be regarded as an electron
doped 11 system with possible local moment magnetism.
\end{abstract}


\maketitle 



The discovery of iron-based compounds, typically represented by
\lofa\cite{lofa_discovery} (1111-type),  \bfa122\cite{bfa_discovery}
(122-type) and
\fs11\cite{fs11_discovery_1,fs11_discovery_2}(11-type), has
triggered enormous enthusiasm in searching for the new high
transition temperature superconductors without
copper\cite{nature07045,PhysRevLett.100.247002,0295-5075-83-1-17002,0295-5075-82-1-17009,0295-5075-83-6-67006}.
Ever since the discovery, density functional studies have been
performed to explore the electronic structure and the pairing
mechanism of the system. Calculations have been performed on
\lofa\cite{0295-5075-83-2-27006,singh_1111,PhysRevLett.101.057010,mazin_1111,cao_1111,PhysRevB.78.033111,PhysRevB.78.224517},
and the ground state is found to be a collinear
anti-ferromagnetism (COL) state. The magnetic ordering was proposed
to be the consequence of the Fermi surface nesting
phenomena\cite{0295-5075-83-2-27006,singh_1111,mazin_1111}  , which
is also present in the \bfa122 parent compound\cite{singh_122}.
The Fermi surface nesting is thus considered closely related with
the superconducting (SC) phenomena since it is suppressed in SC
phase. On the other hand, the $J_1$-$J_2$ Heisenberg model based on
a local moment
picture\cite{PhysRevLett.101.076401,PhysRevLett.101.057010,PhysRevB.78.224517}
was also proposed to account for the magnetic structure. It is also
found that the band energy dispersion of these compounds should be
calculated using the unrelaxed experimental structure in order to
compare with the experiments\cite{njp_11_025023,PhysRevB.81.224503},
and that the ordered magnetic moments on Fe atom are systematically
overestimated in density functional calculations.

Recently, another substance with similar chemical composition,
\kfs122, has been produced\cite{kfs122_discovery}. The material is
reported to be iso-structural to \bfa122, and the superconductivity as
high as 30 K is reported when it is intrinsically doped (\kfsx22
with $x=0.2$). Question thus arises that whether this material
represents a new family or it is one of the discovered class. More
specifically, since \kfs122 is structurally close to \bfa122 but
chemically close to \fs11, it is interesting to clarify which one is
closer to its electronic structure.

In this paper, we report our first-principles study of this
compound. We demonstrate that the electronic structure of the
parent \kfs122 could be regarded as an electron doped 11
system, instead of the structurally much closer \bfa122. All the
calculations were performed with the {\sc Quantum ESPRESSO}
code\cite{QE-2009}, while an accurate set of PAW data\cite{cao_paw}
were employed throughout the calculation. A 48 Ry energy cut-off
ensures the calculations converge to 0.1 mRy, and all structures
were optimized until forces on individual atoms were smaller than
0.1 mRy/bohr and external pressure less than 0.5 kbar. For
non-magnetic (NM) and checkerboard anti-ferromagnetic (CBD) states, a
$8\times8\times8$ Monkhorst-Pack k-grid\cite{mp_kgrid} was found to
be sufficient; while for the collinear anti-ferromagnetic (COL)
state and bi-collinear anti-ferromagnetic (BIC) state, $6\times6\times8$ and $16\times8\times4$ Monkhorst-Pack k-grid were needed to
ensure the convergence to $<$ 1 meV/Fe, respectively. The PBE flavor of general gradient
approximation (GGA) to the exchange-correlation
functional\cite{PBE_xc} was applied throughout the calculations.

We first examine several possible spin configurations for \kfs122
(TABLE \ref{tab:structure}). The column expt indicates calculation
with experimental structure and non-magnetic spin configuration,
while the structures are fully optimized (lattice constants as well
as internal coordinates) for NM/CBD/COL/BIC columns. It is therefore
apparent that the collinear phase, which is 57 meV/Fe lower than the NM
phase, is the ground state of \kfs122. This magnetic state ordering
was also double-checked with full-potential linearized augmented
plane wave (FLAPW) method using the elk code\cite{elk_code}. A
body-centered tetragonal (bct) to base-centered orthorhombic (bco)
phase transition is also present in the process, although the
orthorhombicity $\epsilon=1-b/a=0.03\%$ is almost negligible.
Although the phase transition is not yet observed in the
experiments, the resistivity measurement shows an abrupt change
around $T=100$K\cite{kfs122_discovery}, which we propose to be due
to the bco-bct phase transition. The magnetic moment on Fe atoms
turns out to be 2.26 $\mu_B$/Fe for the collinear phase, which is similar to
those obtained in PAW calculations for \bfa122\cite{cao_paw}. In
order to estimate the magnetic coupling strength, we incorporate the
$J_1$-$J_2$ Heisenberg model, defined by
$$
H=J_1\sum_{<i,j>}{\vec S}_i\cdot{\vec S}_j +J_2\sum_{<<i,j>>}{\vec
S}_i\cdot{\vec S}_j.
$$
Where,  ${\vec S}_i$ is the spin operator (of
 magnitude $S$) at the site $i$, $<i,j>$ and $<<i,j>>$ denote the summation over the nearest
neighbor and the next nearest neighbor sites, $J_1$ and $J_2$ are
the nearest neighbor and the next nearest neighbor exchange
interactions, respectively. By calculating the total energy per Fe
atom for the CBD and COL states and assuming $S=1$,
we obtain
 $J_2=29$ meV and
  $J_1=38$ meV, respectively.

\begin{table}[htp]
 \caption{Geometry, energetic and magnetic properties of \kfs122. Results in column expt were obtained using the experimental structure and spin-unpolarized calculations; while the NM/CBD/COL/BIC columns correspond to non-magnetic/checkerboard AFM/collinear AFM/bi-collinear AFM configurations using the fully optimized structures (lattice constants as well as internal coordinates), respectively. $E_{\Delta}$ is the total energy difference per iron atom referenced to the fully optimized NM structure, and $m_{\mathrm{Fe}}$ is the local magnetic moment on Fe.}
 \label{tab:structure}
 \begin{tabular}{c|c|c|c|c|c}
   & expt & NM & CBD & COL & BIC\\
  \hline
  a (\AA)& 3.9136 & 3.8791 & 3.9058 & 5.5930 & 3.8271 \\
  b (\AA) & 3.9136 & 3.8791 & 3.9058 & 5.5916 & 7.9530 \\
  c (\AA) & 14.0367 & 13.4476 & 13.6849 & 13.8525 & 14.4783\\
  $E_{\Delta}$ (meV/Fe) & 272 & 0 & -18 & {\bf -57} & -39\\
  $m_{\mathrm{Fe}}$ ($\mu_B$) & 0.0 & 0.0 & 1.49 & 2.26 & 2.58\\
 \end{tabular}
\end{table}

\begin{figure}[htp]
 \centering
 \rotatebox{270}{\scalebox{0.6}{\includegraphics{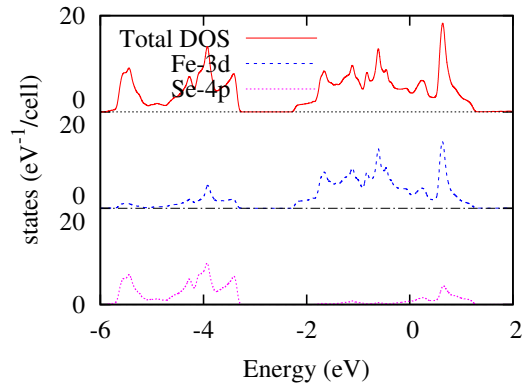}}}
 \caption{Total and projected density of states of \kfs122. The upper panel (solid line) is the total DOS; middle panel (dashed line) is the PDOS on Fe-3d orbitals; lower panel (dotted line) is the PDOS on Se-4p orbitals. We show only the energy range from $E_F-6.0$ eV to $E_F+2.0$ eV.\label{figure:dos}}
\end{figure}

Calculations with \lofa, \bfa122, and \fs11 systems suggest that the band dispersions and DOS of these systems should be studied without structural optimization in order to compare with the experiments\cite{mazin_1111,singh_1111,cao_1111,singh_122,alaska_11,singh_11,singh_review}; although their energetic properties as well as magnetism should be explored with structural relaxation. We followed this procedure, and the discussion in the rest of this paper will primarily focus on the calculations with unrelaxed (experimental) structure unless we explicitly specify. Firstly, we present the density of states (DOS), as well as the projected density of states (PDOS) calculations (FIG. \ref{figure:dos}). The DOS and PDOS of \kfs122 resemble those of \bfa122 systems, and exhibit typical characteristics of the layered structure. The contribution from Fe-3d and Se-4p orbitals dominates the states from $E_F-6.0$ eV to $E_F+2.0$ eV, while most of the K-4s contribution locates from $E_F+2.0$ eV to $E_F+6.0$ eV. A closer examination of the PDOS data shows that over 90\% of the states from $E_F-2.0$ eV to $E_F$ are from the Fe-3d orbitals, and that the Fe-3d/Se-4p orbital hybridizes considerably from $E_F-6.0$ eV to $E_F-3.2$ eV and from $E_F$ to $E_F+2.0$ eV.

\begin{figure}[htp]
 \centering
 \rotatebox{270}{\scalebox{0.6}{\includegraphics{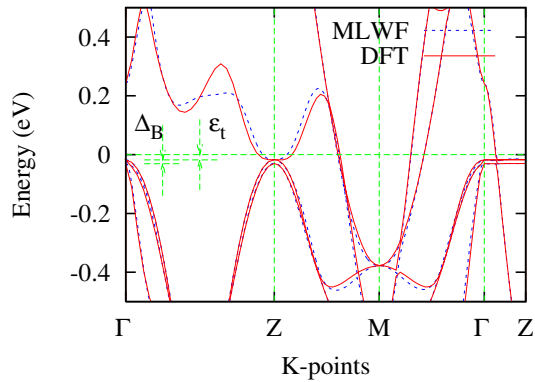}}}
 \caption{Band structure of non-magnetic \kfs122 calculated with the experimental structure. The solid line is the density functional theory (DFT) result and the dashed line is fitted with the maximally localized wannier function (MLWF) method.\label{figure:bs}}
\end{figure}

We further calculated the band structure for \kfs122, as shown in
FIG. \ref{figure:bs}. Since the Se atom (4s$^2$4p$^4$) outermost shell
has 1 more electron than the As atom (4s$^2$4p$^3$), the FeSe layer
could be regarded as highly electron-doped. In fact, the band
structure of \kfs122 shows that the Fe-3d$_{zx(y)}$ and
Fe-3d$_{x^2-y^2}$ bands are fully occupied, whereas these bands were
the origin of the hole pockets in the \bfa122 systems. At $\Gamma$,
the 3d$_{zx}$ and 3d$_{zy}$ bands becomes degenerate due to the
crystal symmetry. We define the energy difference between these two
bands and the Fermi level at $\Gamma$ to be
$\epsilon_t=E_F-\epsilon_{\Gamma}^{zx(y)}$, and the difference
between these two bands and the 3d$_{x^2-y^2}$ band at $\Gamma$ to be
$\Delta_B=\epsilon_{\Gamma}^{zx(y)}-\epsilon_{\Gamma}^{x^2-y^2}$
(FIG. \ref{figure:bs}). The latter is due to the slightly deformed
tetrahedral crystal field by the 4 Se atoms around the Fe atom. For
the \kfs122 systems, $\epsilon_t$ and $\Delta_B$ are 18 meV and 13
meV, respectively; while for the \bfa122 systems, they are -297 meV
and 204 meV, respectively. Similar to the \bfa122 band structure,
the bands close to $E_F$ from $\Gamma$ to $Z$ are mostly flat,
except for the one cross the Fermi level which is due to the
hybridization of Fe-3d and Se-4p orbitals. It is worthy noting that
the structural relaxation does not change the number of bands across
the Fermi level, nor the orbital character of these bands for
\kfs122, in contrast to the cases for \lofa\ and \bfa122. However,
the structural optimization expands the band splittings $\Delta_B$
to 314 meV, and shifts the top of fully occupied d bands
$\epsilon_t$ to 149 meV.

The band structure we obtained is then fitted using the maximally
localized Wannier function (MLWF) method \cite{mlwf_1,mlwf_2} (FIG.
\ref{figure:bs}) to obtain a model Hamiltonian for reconstructing
the Fermi surfaces. To perform the fitting, we chose the 16 bands
from $E_F-6.0$ eV to $E_F+2.0$ eV, and 16 initial guess orbitals
including the Fe-3d and Se-4p to ensure the fitting quality.
Nevertheless, it is possible to fit the band structure with slightly
worse quality with the 10 Fe-3d orbitals only, in order to reduce
the Hamiltonian size. Applying the symmetry, the number of orbitals
could be further brought down to five.

\begin{figure}[htp]
 \centering
 \subfigure[$x=0.0$]{
  \scalebox{0.36}{\includegraphics{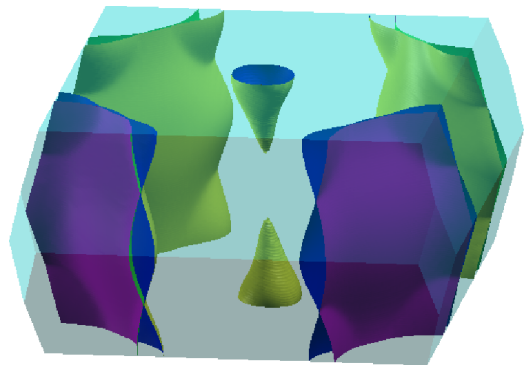}}
  \label{figure:fs_0.0}
 }
 \subfigure[$x=0.2$]{
  \scalebox{0.36}{\includegraphics{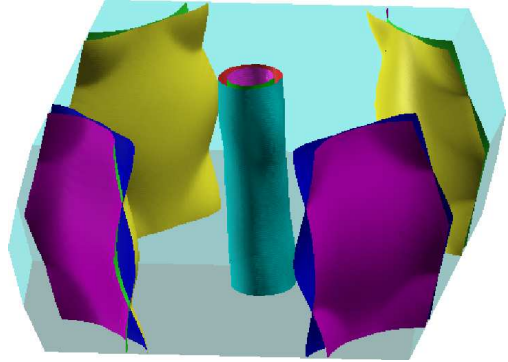}}
  \label{figure:fs_0.2}
 }
 \subfigure[$x=0.0$]{
  \rotatebox{270}{\scalebox{0.52}{\includegraphics{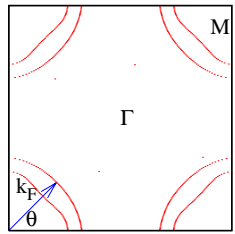}}}
  \label{figure:fs_z0}
 }
 \subfigure[]{
  \rotatebox{270}{\scalebox{0.52}{\includegraphics{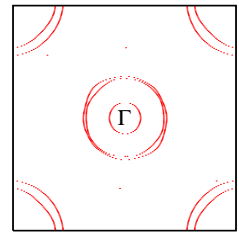}}}
  \label{figure:fs_z1}
 }
 \caption{Fermi surface of \kfsx22 reconstructed using the MLWFs. Panel \ref{figure:fs_0.0} and \ref{figure:fs_z0} are the plots for the parent compound ($x=0.0$); panel \ref{figure:fs_0.2} is the plot for $x=0.2$; panel \ref{figure:fs_z1} is obtained by shifting $E_F$ by -0.2 eV. Panel \ref{figure:fs_z0} and \ref{figure:fs_z1} are the 2-D plots of cross-section at $k_z=0.0$. Only the $k_F$ vector of the outer sheet is drawn in Fig. \ref{figure:fs_z0} for the sake of visibility.\label{figure:fs}}
\end{figure}

We show the parent \kfs122 Fermi surface in FIG.
\ref{figure:fs_0.0}. The Fermi surface of the parent \kfs122 consists
of two sheets around $M$-point and one sheet around $\Gamma$.
Despite of the similarities between the \kfs122 and the \bfa122
systems, two distinctions are apparent. First of all, the two sheets
around $M$ points in \kfs122 are much more symmetric than those in
\bfa122. If we define $k_F(\theta)$ to be the k-vector from the $M$
point to the Fermi surface sheet, where $\theta$ denotes the angle
formed by the vector and $M$-$M'$ (as shown in FIG.
\ref{figure:fs_z0}), we could further define
$\eta=1-\frac{k_F(45^{\circ})}{k_F(-45^{\circ})}$. For the two
sheets around $M$ in \kfs122, both yield $\eta<1\%$, while for the
\bfa122 system the inner sheet has $\eta=48\%$ and the outer sheet
has $\eta=41\%$. This feature suggests that the electronic structure
of \kfs122 is in fact much closer to \fs11 instead of \bfa122.
Secondly, the sheets around $\Gamma$ are completely different
between \kfs122 and \bfa122 or \fs11. Three sheets were observed for
\bfa122 or \fs11 system, which constitutes the three hole pockets
for the system. For \kfs122 system, only one sheet exists around
(0,0,$k_z$) axis, which is highly 3-D and vanishes around $\Gamma$.
Thus, the Fermi surface nesting effect is absent in the parent
\kfs122 compound. Nevertheless, one could achieve \fs11-like fermi
surface using the rigid band model simply by shifting down the fermi
level (FIG. \ref{figure:fs_z1}), or effectively by hole doping. From
the band structure calculation, the Fermi level has to be shifted
down by 0.2 eV in order to recover the Fermi surface nesting effect.
The DOS result indicates that shifting down $E_F$ by 0.2 eV
corresponds to 1.0 $\vert e\vert$ hole doping effectively, or
completely removing K from \kfs122. Due to the loss of Fermi surface
nesting, the magnetism of \kfs122 is not simply due to the Fermi
surface nesting effect. Instead, it is possible that the localized
Fe-3d orbitals plays an essential role in the magnetism. Using the
same model, we could also obtain the Fermi surface for \kfsx22
($x=0.2$), as shown in FIG. \ref{figure:fs_0.2}, which could be an
analogue to an electron-doped 11 system.


Finally, we study the $U$-dependence of the magnetic coupling strength $J_1$ and $J_2$ using the GGA+$U$ method, to test if the COL configuration remains as the ground state if there is electron correlation beyond LDA. A series of $U$ from 1.0 eV to 6.0 eV were used to optimize the lattice constants as well as the internal coordinates for NM/CBD/COL configurations, and then $J_1$ and $J_2$ under different $U$ were fitted using $m_{\mathrm{Fe}}=1.0 \mu_B$. Both $J_1$ and $J_2$ show linear dependence with respect to the on-site energy $U$, and the collinear state remains the ground state within a reasonable $U$ range.

In conclusion, we have studied the electronic structure of \kfs122
using first-principles calculations. The ground state of \kfs122
turns out to be collinear anti-ferromagnetic configuration with
2.26 $\mu_B$ magnetic moment on Fe atoms, similar to \bfa122. The
$J_1$ and $J_2$ coupling strengths are calculated to be 0.038 eV and
0.029 eV, respectively. Although the band structure is similar to
heavily electron-doped \bfa122, the Fermi surface suggests that the
system is much closer to an electron-doped \fs11 system. The
Fermi surface nesting effect is absent in the parent \kfs122
compound, thus the antiferromagnetism is possibly due to the local
moments instead of the itinerant electrons. Since the \kfs122 is
electron-doped, the superconductivity could be introduced with
hole-doping or, effectively, K or Fe deficiencies.

We would like to thank Hong Ding and Gang Wang for calling our
attention to the \kfs122 compound reported in Ref.\cite{kfs122_discovery}. We also thank
Guanghan Cao, Xiaoyong Feng, and Qimiao Si for helpful discussions.
This work was supported by the NSFC, the 973 Project of the MOST and
the Fundamental Research Funds for the Central Universities of China
(No. 2010QNA3026). All the calculations were performed on Hangzhou
Normal University College of Science High Performance Computing
Center.

Note added: After submitting this work to arXiv, we became aware of
two recent papers\cite{arxiv:1012.5536,arxiv:1012.5164} where some
related calculations have been also performed on the KFe$_2$Se$_2$
compound.

\bibliography{k122}
\end{document}